\documentclass[preprints,article,accept,pdftex,moreauthors]{Definitions/mdpi} 

\firstpage{1} 
\pubvolume{1}
\issuenum{1}
\articlenumber{0}
\pubyear{2024}
\copyrightyear{2024}
\datereceived{ } 
\daterevised{ } 
\dateaccepted{ } 
\datepublished{ } 
\hreflink{https://doi.org/} 

\usepackage{subcaption}

\allowdisplaybreaks[1]

\Title{A Generic Analysis of Nucleon Decay Branching  Fractions in Flipped SU(5) Grand Unification}

\TitleCitation{A Generic Analysis of Nucleon Decay Branching  Fractions in Flipped SU(5) Grand Unification}

\Author{Koichi Hamaguchi $^{1,2}$\orcidA{}, Shihwen Hor $^{1}$\orcidB{}, Natsumi Nagata $^{1}$*\orcidC{}, and Hiroki Takahashi $^{1}$}

\AuthorNames{Firstname Lastname, Firstname Lastname and Firstname Lastname}

\AuthorCitation{Lastname, F.; Lastname, F.; Lastname, F.}

\address{%
$^{1}$ \quad Department of Physics, University of Tokyo, Bunkyo-ku, Tokyo 113--0033, Japan\\
$^{2}$ \quad Kavli Institute for the Physics and Mathematics of the Universe (Kavli IPMU), University of Tokyo, Kashiwa, Chiba 277--8583, Japan}

\corres{Correspondence: \href{mailto:natsumi@hep-th.phys.s.u-tokyo.ac.jp}{\tt natsumi@hep-th.phys.s.u-tokyo.ac.jp}}

\abstract{
In flipped SU(5) grand unified theories, the partial decay lifetimes of certain nucleon decay channels depend generically on an unknown unitary matrix, which arises when left-handed lepton fields are embedded into anti-fundamental representations of SU(5). This dependency is particularly relevant when the neutrino mass matrix has a generic structure, introducing uncertainty in the prediction of nucleon decay branching fractions within flipped SU(5). In this paper, we demonstrate that this uncertainty can be parametrized using two parameters, which can be determined by measuring the partial lifetimes of $p \to \pi^0 e^+$, $p \to \pi^0 \mu^+$, and $n \to \pi^0 \bar{\nu}$. In addition, we establish upper limits on the ratios of the decay widths of these channels, offering a potential method to test flipped SU(5) in future nucleon decay experiments.}


\begin{document}

\section{Introduction}

Flipped $\mathrm{SU}(5) \times \mathrm{U}(1)$ grand unified theories (GUTs)~\cite{Barr:1981qv, Derendinger:1983aj, Antoniadis:1987dx, Barr:1988yj} offer a promising framework for constructing GUT models. This framework includes right-handed neutrinos as an essential component and naturally explains small neutrino masses via the seesaw mechanism~\cite{Minkowski:1977sc, Yanagida:1979as, Glashow:1979nm, Gell-Mann:1979vob, Mohapatra:1979ia}, in contrast to the standard SU(5)~\cite{Georgi:1974sy}, where right-handed neutrinos are not required, as they are singlets of SU(5) and do not complete the SU(5) representations. In addition, flipped SU(5) incorporates the missing-partner mechanism~\cite{Masiero:1982fe, Grinstein:1982um, Antoniadis:1987dx} to naturally solve the doublet-triplet splitting problem.

Flipped SU(5) predicts a distinctive pattern of nucleon decay branching fractions~\cite{Ellis:1993ks, Ellis:2002vk, Dorsner:2004xx, Li:2010dp, Ellis:2020qad, Hamaguchi:2020tet, Mehmood:2020irm, Haba:2021rzs, Ellis:2021vpp, Mehmood:2023gmm, King:2023wkm}. The missing-partner mechanism suppresses the contribution of color-triplet Higgs exchange~\cite{Ellis:1988tx} to dimension-five operators~\cite{Weinberg:1981wj, Sakai:1981pk}, which contrasts sharply with the minimal supersymmetric SU(5)~\cite{Dimopoulos:1981zb, Sakai:1981gr}, where dimension-five proton decay operators could be problematic~\cite{Goto:1998qg, Murayama:2001ur} unless the supersymmetric particle mass scale is very high~\cite{McKeen:2013dma, Liu:2013ula, Hisano:2013exa, Nagata:2013sba, Nagata:2013ive, Evans:2015bxa, Bajc:2015ita, Ellis:2015rya, Ellis:2016tjc, Ellis:2017djk, Evans:2019oyw, Ellis:2019fwf, Ellis:2020mno, Hamaguchi:2022yjo, Kim:2024bub}. Its contribution to dimension-six operators is also small~\cite{Ellis:1988tx} if the color-triplet Higgs lies around the GUT scale. Consequently, the SU(5) gauge boson exchange predominantly drives nucleon decays in flipped SU(5). Since this process is induced by gauge interactions, it allows for relatively robust predictions of certain nucleon decay branching fractions. For instance, the decay width of $p \to K^+ \bar{\nu}$ is predicted to vanish due to the unitarity of the Cabibbo–Kobayashi–Maskawa (CKM) matrix~\cite{Ellis:1993ks}. Ratios of decay widths, such as $\Gamma(p \to K^0 e^+)/\Gamma(p \to \pi^0 e^+)$ and $\Gamma(p \to K^0 \mu^+)/\Gamma(p \to \pi^0 \mu^+)$, are determined by the CKM matrix elements~\cite{Ellis:2020qad}, along with the relevant hadron masses and matrix elements.

However, some important decay channels suffer from uncertainties in their decay branching fraction calculations, even though these processes are induced by gauge interactions. This uncertainty arises from an unknown unitary matrix $U_\ell$, introduced when embedding left-handed lepton fields into anti-fundamental representations of SU(5)~\cite{Ellis:1993ks}. As discussed in Ref.~\cite{Ellis:2020qad}, in some specific scenarios as in Refs.~\cite{Ellis:2019jha, Ellis:2019opr, Ellis:2020lnc}, this unitary matrix corresponds to the Pontecorvo–Maki–Nakagawa–Sakata (PMNS) matrix,  which allows us to make a robust prediction. Generically speaking, however, unknown matrix elements, $(U_\ell)_{11}$ and $(U_\ell)_{12}$, appear in the expressions for the decay widths, causing uncertainty in the calculation. It is therefore crucial to study the impact of this uncertainty on the predicted nucleon decay branching fractions and to evaluate whether flipped SU(5) can still be tested in upcoming nucleon decay experiments~\cite{Dev:2022jbf}.

In this paper, we demonstrate that this uncertainty can be parametrized using only two parameters and these parameters can be determined by measuring the lifetimes of the three nucleon decay channels, $p \to \pi^0 e^+$, $p \to \pi^0 \mu^+$, and $n \to \pi^0 \bar{\nu}$. We also derive upper limits on the ratios between the decay widths of these channels, providing a means to test flipped SU(5) in future experiments.

\section{Nucleon decay in flipped SU(5)}

We first describe the flipped SU(5) model considered in this paper. This model is a supersymmetric (SUSY) $\mathrm{SU} (5) \times \mathrm{U} (1)$ gauge theory, where the three generations of matter fields in the minimal supersymmetric Standard Model (MSSM), together with three right-handed neutrinos, are embedded into $\mathbf{10}$, $\bar{\mathbf{5}}$, and $\mathbf{1}$ representations of SU(5), which we denote by $F_i$, $\bar{f}_i$ and $\ell^c_i$, respectively, with $i=1,2,3$ the generation index. Their U(1) charges in units of $1/\sqrt{40}$ are $+1$, $-3$, and $+5$, respectively. We also add singlet fields $\phi_i$ to give masses to the right-handed neutrinos after the $\mathrm{SU} (5) \times \mathrm{U} (1)$ gauge symmetry is spontaneously broken. This symmetry breaking is triggered by the vacuum expectation values (VEVs) of a pair of $\mathbf{10}$ and $\overline{\mathbf{10}}$ Higgs fields, $H$ and $\bar{H}$, whose U(1) charges are $+1$ and $-1$, respectively. The MSSM Higgs doublet fields, $H_u$ and $H_d$ are embedded into anti-fundamental and fundamental representations of SU(5) with U(1) charges $+2$ and $-2$; these representations are denoted by $\bar{h}$ and $h$, respectively. We further assume that this model possesses two $\mathbb{Z}_2$ symmetries. One is the standard $R$-parity, under which $F_i$, $\bar{f}_i$, $\ell^c_i$, and $\phi_i$ are odd and $H$, $\bar{H}$, $h$, and $\bar{h}$ are even. Another one is to remove the mass term for $H$ and $\bar{H}$, where only the $H$ field is odd and the other fields are even;\footnote{We assume that this symmetry is violated by some
Planck-scale suppressed operators in order to prevent the formation of stable domain walls. } the absence of this mass term is advantageous for suppressing the dimension-five nucleon-decay operators induced by the color-triplet Higgs exchange.  We summarize the field content and charge assignments in Table~\ref{tab:model}.

\begin{table}
    \caption{
        The field content and the charge assignments in the flipped SU(5) model. The U(1) charges are given in units of $1/\sqrt{40}$. 
       }
    \newcolumntype{C}{>{\centering\arraybackslash}X}
    \begin{tabularx}{\textwidth}{l l CCCC}
        \toprule
    Fields & Components & SU(5)\quad & U(1)\quad &  $\mathbb{Z}_2$\quad &  $R$-parity \\
    \midrule
    $F_i$ & $d^c_i$, $Q_i$, $\nu^c_i$ & $\mathbf{10}$ & $+1$ & $+$ & $-$ \\ 
    $\bar{f}_i$ & $u^c_i$, $L_i$ & $\overline{\mathbf{5}}$ & $-3$ &$+$ & $-$\\ 
    ${\ell}_i^c$ & $e^c_i$ & ${\mathbf{1}}$ & $+5$ &$+$ & $-$\\ 
    $\phi_i$ & $\phi_i$ & ${\mathbf{1}}$ & $0$ &$+$ & $-$\\ 
    $H$ & $d^c_H$, $Q_H$, $\nu^c_H$ & $\mathbf{10}$ & $+1$ &$-$ & $+$\\ 
    $\bar{H}$ & ${d}^c_{\bar{H}}$, ${Q}_{\bar{H}}$, ${\nu}^c_{\bar{H}}$ & $\overline{\mathbf{10}}$ & $-1$  &$+$ & $+$\\ 
    $h$ & $D$, $H_d$ & $\mathbf{5}$ & $-2$ &$+$ & $+$\\ 
    $\bar{h}$ & $\bar{D}$, $H_u$ & $\overline{\mathbf{5}}$ & $+2$ &$+$ & $+$\\ 
    \bottomrule
    \end{tabularx}
    \label{tab:model}
    \end{table}

The renormalizable superpotential in this model is given by
\begin{align} 
W &=  \lambda_1^{ij} F_iF_jh + \lambda_2^{ij} F_i\bar{f}_j\bar{h} +
 \lambda_3^{ij}\bar{f}_i\ell^c_j h + \lambda_4 HHh + \lambda_5
 \bar{H}\bar{H}\bar{h}
+ \lambda_6^{ij} F_i\bar{H}\phi_j 
+ \mu^{ij}\phi_i\phi_j ~,
\label{Wgen} 
\end{align}
where we assume the $\mu$-term for $h$ and $\bar{h}$ to be absent\footnote{It is possible to construct a flipped SU(5) model where this $\mu$-term is forbidden by an $R$-symmetry~\cite{Hamaguchi:2020tet}. } so that the doublet-triplet splitting problem can be solved by the missing-partner mechanism. 

The scalar potential of this model has an $F$- and $D$-flat direction along $|\nu_H^c| = |\nu_{\bar{H}}^c|$ and we assume that these fields develop VEVs in this direction to break the $\mathrm{SU} (5) \times \mathrm{U} (1)$ gauge symmetry---the origin of this potential is destabilized by a soft SUSY-breaking mass term and the flat direction at large field values is lifted by a non-renormalizable superpotential term. See Refs.~\cite{Ellis:2017jcp, Ellis:2018moe} for more detailed discussions on this symmetry breaking. After $H$ and $\bar{H}$ acquire VEVs, the color components in these fields, $d_H^c$ and $d^c_{\bar{H}}$, form vector-like mass terms with those in $h$ and $\bar{h}$, $D$ and $\bar{D}$, via the couplings $\lambda_4$ and $\lambda_5$ in Eq.~\eqref{Wgen}, while the MSSM doublet Higgs fields remain massless---\textit{i.e.}, the doublet-triplet splitting is realized. $Q_H$, $Q_{\bar{H}}$, and a linear combination of $\nu_H^c$ and $\nu_{\bar{H}}^c$ are eaten by the gauge fields corresponding to the broken symmetries, giving them masses of the order of the GUT scale. The other combination of $\nu_H^c$ and $\nu_{\bar{H}}^c$ acquires a mass of the order of the soft SUSY-breaking scale. The VEVs of $H$ and $\bar{H}$ also generate mass terms for the right-handed neutrinos $\nu_i^c$ and the singlet fields $\phi_i$ via the couplings $\lambda_6^{ij}$. The integration of these fields then gives rise to the dimension-five Weinberg operators~\cite{Weinberg:1979sa}, which generate neutrino masses after $H_u$ acquires a VEV. 

To see the flavor structure of this model, we take a flavor basis where $\lambda_2^{ij}$ is real and diagonal without loss of generality. In this case, we find that the MSSM matter fields and right-handed neutrinos are embedded into $F_i$, $\bar{f}_i$, and $\ell^c_i$ as 
\begin{align}
    F_i &= \left\{Q_i, ~ V_{ij} e^{-i \varphi_j} d_j^c, ~ \left(U_{\nu^c}\right)_{ij}
    \nu^c_j \right\} ~, \nonumber \\
    \bar{f}_i &= \left\{
   u_i^c ~, \left(U_\ell^* \right)_{ij} L_j 
   \right\} ~, \nonumber \\
    \ell^c_i &= \left(U_{\ell^c}\right)_{ij} e_j^c ~,
   \label{eq:embedding}
\end{align}
where $V$ is the CKM matrix, $U_{\nu^c}$, $U_\ell$, and $U_{\ell^c}$ are unitary matrices, and $\varphi_i$ are real phase parameters satisfying the condition $\sum_{i} \varphi_i = 0$. For the detailed procedure to obtain this result, see Refs.~\cite{Ellis:1993ks, Hamaguchi:2020tet}. The doublet quark and lepton fields are expressed in terms of the mass eigenstates as 
\begin{equation}
 Q_i = 
\begin{pmatrix}
 u_i \\ V_{ij} d_j
\end{pmatrix}
~, \qquad
L_i = 
\begin{pmatrix}
 U_{ij} \nu_j \\ e_i
\end{pmatrix}
~,
\end{equation}
where $U$ is the PMNS matrix. We also need another unitary matrix $U_\nu$ to diagonalize the mass matrix of light neutrinos $m_\nu$: 
\begin{equation}
    m_\nu^D = U_\nu^T m_\nu U_\nu ~. 
    \label{eq:mnud}
\end{equation}
We find that the PMNS matrix is related to the unitary matrices $U_\ell$ and $U_\nu$ as 
\begin{equation}
    U = U^T_\ell U_\nu ~.
    \label{eq:uellunu}
\end{equation}
Notice that if the neutrino mass matrix $m_\nu$ is predicted, as in the case of the scenario considered in Refs.~\cite{Ellis:2019jha, Ellis:2019opr, Ellis:2020lnc}, $U_\nu$ is obtained from Eq.~\eqref{eq:mnud} and then $U_\ell$ is given by Eq.~\eqref{eq:uellunu} as a function of the PMNS matrix. Otherwise, we have ambiguity in the determination of $U_\ell$ and $U_\nu$. 

Now we discuss the nucleon decay in this model. The exchange of the SU(5) gauge bosons generates the effective operators of the form 
\begin{equation}
    {\cal L}_{\rm eff}
   =C^{ijkl}{\cal O}_{ijkl}
   + \mathrm{h.c.} 
   ~,
   \label{eq:l6eff}
   \end{equation}
   where 
   \begin{align}
    {\cal O}_{ijkl}&\equiv \int d^2\theta d^2\bar{\theta}~
   \epsilon_{abc}\epsilon_{\alpha\beta}
   \bigl(u^{c\dagger}_i\bigr)^a
   \bigl(d^{c\dagger} _j\bigr)^b
   e^{-\frac{2}{3}g^\prime B}
   \bigl(e^{2g_3G}Q_k^\alpha\bigr)^cL^\beta_l~,
\end{align}
with $G$ and $B$ the SU(3)$_C$ and U(1)$_Y$ gauge vector superfields, respectively, and $g_3$ and $g^\prime$ the corresponding gauge couplings. Notice that another type of the dimension-six operators allowed by the Standard Model gauge symmetry is not induced in flipped SU(5), as the right-handed charged lepton fields $e^c_i$ are singlets of SU(5). We also note that as the interaction~\eqref{eq:l6eff} is induced by the gauge interaction, the following analysis is applicable not only to the SUSY flipped SU(5) model but also to non-SUSY ones. The Wilson coefficients $C^{ijkl}$ are given by 
\begin{equation}
    C^{ijkl} = \frac{g_5^2}{M_X^2} (U_\ell^*)_{il} V_{kj}^* e^{i\varphi_j} ~,
\end{equation}
where $M_X$ is the mass of the SU(5) gauge bosons and $g_5$ is the SU(5) gauge coupling. 

By using the effective interactions in Eq.~\eqref{eq:l6eff}, we compute the nucleon decay widths, with taking the renormalization-group effect into account. For the detailed discussion on this prescription, see Ref.~\cite{Ellis:2020qad}. In what follows, we summarize the resultant expressions given in Ref.~\cite{Ellis:2020qad}. 

Let us begin with the $p \to \pi^0 \ell_i^+$ decay channels, where $\ell_1^+$ and $\ell_2^+$ denote positron and anti-muon, respectively. The decay widths of these channels are expressed as 
\begin{align}
    \Gamma (p \to \pi^0 \ell_i^+) = \frac{g_5^4 m_p |V_{ud}|^2 |(U_\ell)_{1i}|^2}{32\pi M_X^4} \biggl(1 - \frac{m_\pi^2}{m_p^2}\biggr) ^2 A^2 \left( \langle \pi^0 |(ud)_R u_L |p\rangle_{\ell_i} \right)^2 ~,
    \label{eq:ptopil}
\end{align}
where $m_p$ and $m_\pi$ denote the masses of proton and pion,
respectively; $A$ is a renormalization factor; the quantity in the last parenthesis is the hadron matrix element, for which we use the results obtained from the QCD lattice simulation in Ref.~\cite{Yoo:2021gql}: 
\begin{align}
    \langle \pi^0 |(ud)_R u_L |p\rangle_{e} &= \frac{1}{\sqrt{2}} \langle \pi^+ |(ud)_R d_L |p\rangle_{e} =  - 0.112~\mathrm{GeV}^2 ~, \nonumber \\ 
    \langle \pi^0 |(ud)_R u_L |p\rangle_{\mu} &= \frac{1}{\sqrt{2}} \langle \pi^+ |(ud)_R d_L |p\rangle_{\mu} =  - 0.114~\mathrm{GeV}^2 ~. 
    \label{eq:pi0pipl}
\end{align}
Generically, the current precision of the calculation of the hadron matrix elements is $\mathcal{O} (10)$\%~\cite{Yoo:2021gql}. As we see, the decay widths~\eqref{eq:ptopil} depend on the matrix elements $(U_\ell)_{1i}$. 

There are other decay channels which are related to the above channels. For example, the neutron decay channels $n \to \pi^- \ell_i^+$ are simply related to the above channels via SU(2) isospin relations as 
\begin{equation}
    \Gamma (n \to \pi^- \ell_i^+) = 2 \Gamma (p \to \pi^0 \ell_i^+) ~.
\end{equation}
We note that the current limits on the $n \to \pi^- \ell_i^+$ decay lifetimes are~\cite{Super-Kamiokande:2017gev}
\begin{align}
    \tau (n \to  \pi^- e^+) &> 5.3 \times 10^{33} ~ \mathrm{years} ~, \\ 
    \tau (n \to \pi^- \mu^+) &> 3.5 \times 10^{33} ~ \mathrm{years} ~,
\end{align}
which are weaker than the corresponding proton decay channels by more than a factor of two~\cite{Super-Kamiokande:2020wjk}: 
\begin{align}
    \tau (p \to  \pi^0 e^+) &> 2.4 \times 10^{34} ~ \mathrm{years} ~, \\ 
    \tau (p \to \pi^0 \mu^+) &> 1.6 \times 10^{34} ~ \mathrm{years} ~.
\end{align}
Namely, the $p \to \pi^0 \ell_i^+$ channels impose stronger constraints than the the $n \to \pi^- \ell_i^+$ channels.

The decay channels containing $\pi^0$ in the final state is related to those with $\eta$ as 
\begin{align}
    \frac{\Gamma (p \to \eta \ell_i^+)}{\Gamma (p \to \pi^0 \ell_i^+)} = \frac{(1 - m_\eta^2/m_p^2)^2}{(1-m_\pi^2/m_p^2)^2} \frac{\left( \langle \eta |(ud)_R u_L |p\rangle_{\ell_i} \right)^2}{\left( \langle \pi^0 |(ud)_R u_L |p\rangle_{\ell_i} \right)^2} 
    ~.
    \label{eq:etamodes}
\end{align}
Notice that these ratios are independent of unknown parameters such as $M_X$, $g_5$,  $A$,  and $(U_\ell)_{1i}$. These ratios are found to be numerically small. For example, for the positron mode,\footnote{We do not show the value for the $\mu^+$ channel, since, as stated in Ref.~\cite{Yoo:2021gql}, the treatment of the matrix elements for $\mu^+$ in Ref.~\cite{Aoki:2017puj} is inappropriate. In any case, we expect that the difference from the $e^+$ case is $\mathcal{O} (m_\mu/m_p)$, with $m_\mu$ the muon mass, and thus it does not affect our conclusion.  } the hadron matrix element is computed as~\cite{Aoki:2017puj}
\begin{equation}
    \langle \eta |(ud)_R u_L |p\rangle_{e} = 0.006~\mathrm{GeV}^2 ~,
\end{equation} 
with which the ratio is estimated to be as small as $1.3 \times 10^{-3}$. The current limits on the $\eta$ channels are~\cite{Super-Kamiokande:2017gev} 
\begin{align}
    \tau (p \to  \eta e^+) &> 1.0 \times 10^{34} ~ \mathrm{years} ~, \\ 
    \tau (p \to \eta \mu^+) &> 4.7 \times 10^{33} ~ \mathrm{years} ~,
\end{align}
and thus the $\eta$ channels give much weaker constraints on flipped SU(5) than the pion channels. 

For the proton decay channels including a neutral kaon in the final state, their decay widths are given by 
\begin{align}
    \Gamma (p \to K^0 \ell_i^+) = \frac{g_5^4 m_p |V_{us}|^2 |(U_\ell)_{1i}|^2}{32\pi M_X^4} \biggl(1 - \frac{m_K^2}{m_p^2}\biggr) ^2 A^2 \left( \langle K^0 |(us)_R u_L |p\rangle_{\ell_i} \right)^2 ~,
\end{align}
where $m_K$ is the kaon mass and~\cite{Yoo:2021gql} 
\begin{align}
    \langle K^0 |(us)_R u_L |p\rangle_{e} &=  0.0854~\mathrm{GeV}^2 ~, \nonumber \\ 
    \langle K^0 |(us)_R u_L |p\rangle_{\mu} &= 0.0860~\mathrm{GeV}^2 ~. 
\end{align}
Again, we can remove the dependence on unknown parameters by taking the ratios, 
\begin{equation}
    \frac{\Gamma (p \to K^0 \ell_i^+)}{\Gamma (p \to \pi^0 \ell_i^+)} = \frac{|V_{us}|^2}{|V_{ud}|^2} \frac{(1 - m_K^2/m_p^2)^2}{(1-m_\pi^2/m_p^2)^2} \frac{\left( \langle K^0 |(us)_R u_L |p\rangle_{\ell_i} \right)^2}{\left( \langle \pi^0 |(ud)_R u_L |p\rangle_{\ell_i} \right)^2} \simeq 0.02 ~.
\end{equation}
The current experimental limits on these channels are~\cite{Super-Kamiokande:2005lev, Super-Kamiokande:2022egr} 
\begin{align}
    \tau (p \to  K^0 e^+) &> 1.0 \times 10^{33} ~ \mathrm{years} ~, \\ 
    \tau (p \to K^0 \mu^+) &> 3.6 \times 10^{33} ~ \mathrm{years} ~,
\end{align}
which are much weaker than the limits on the $\pi^0$ channels. We note in passing that the effective Lagrangian in Eq.~\eqref{eq:l6eff} does not induce $n \to K^- \ell_i^+$. 

All in all, for the $\ell_i^+$ channels, $p \to \pi^0 e^+$ and $p \to \pi^0 \mu^+$ provide most sensitive probes of flipped SU(5). The decay widths of these channels depend on the unknown matrix elements $(U_{\ell})_{11}$ and $(U_{\ell})_{12}$, respectively. The decay rates of the other $\ell_i^+$ channels are related to these two channels through the multiplication of constant factors, as we described above. 

Next, we discuss the anti-neutrino channels. As nucleon-decay experiments are unable to detect the anti-neutrino in the final state, we take the sum over all the neutrino generations when we calculate the decay widths. For the $p \to \pi^+ \bar{\nu}$ channel, we have 
\begin{equation}
   \Gamma (p \to \pi^+ \bar{\nu}) \equiv  \sum_i \Gamma (p \to \pi^+ \bar{\nu}_i) =   \frac{g_5^4 m_p }{32\pi M_X^4} \biggl(1 - \frac{m_\pi^2}{m_p^2}\biggr) ^2 A^2 \left( \langle \pi^+ |(ud)_R d_L |p\rangle_e \right)^2 ~.
\end{equation}
As we see, this expression does not depend on $U_\ell$; this follows from the unitarity condition $\sum_i |(U_\ell)_{1i}|^2 = 1$. The $n \to \pi^0 \bar{\nu}$ channel is related to this channel by isospin and we find 
\begin{equation}
    \Gamma (n \to \pi^0 \bar{\nu}) = \frac{1}{2} \Gamma (p \to \pi^+ \bar{\nu}) ~. 
\end{equation}
The current experimental constraints on these channels are~\cite{Super-Kamiokande:2013rwg} 
\begin{align}
    \tau (p \to  \pi^+ \bar{\nu}) &> 3.9 \times 10^{32} ~ \mathrm{years} ~, \\ 
    \tau (n \to \pi^0 \bar{\nu}) &> 1.1 \times 10^{33} ~ \mathrm{years} ~.
\end{align}
We see that in this case the limit on the neutron decay channel is better than that on the proton decay channel by more than a factor of two. We thus consider $n \to \pi^0 \bar{\nu}$, rather than $p \to  \pi^+ \bar{\nu}$, in the following analysis. 

Finally, it follows from the unitarity of the CKM matrix that the kaon channels are not induced by the interaction in Eq.~\eqref{eq:l6eff}~\cite{Ellis:1993ks}: 
\begin{equation}
    \Gamma (p \to K^+ \bar{\nu}) = \Gamma (n \to K^0 \bar{\nu}) = 0 ~.
\end{equation}
This is a characteristic prediction of flipped SU(5), and, in particular, the detection of these modes excludes flipped SU(5). This is in stark contrast to the case of the minimal SUSY SU(5), where $p \to K^+ \bar{\nu}$ induced by the exchange of the color-triplet Higgs tends to be the dominant decay channel~\cite{Dimopoulos:1981dw, Ellis:1981tv}.

In conclusion, to examine the prediction of flipped SU(5) in future nucleon-decay experiments, the $p \to \pi^0 e^+$, $p \to \pi^0 \mu^+$, and $n \to \pi^0 \bar{\nu}$ channels are most useful. We will discuss in the subsequent section that we can determine the unknown matrix elements $|(U_\ell)_{11}|$ and $|(U_\ell)_{12}|$ by measuring the partial decay widths of these channels.

\section{Results}

In what follows, we consider the ratios
\begin{align}
    R_{\ell_i} \equiv \frac{\Gamma (p \to \pi^0 \ell_i^+)}{\Gamma (n \to \pi^0 \bar{\nu})}  
    = 2 |V_{ud}|^2 |(U_\ell)_{1i}|^2 \frac{\left( \langle \pi^0 |(ud)_R u_L |p\rangle_{\ell_i} \right)^2}{\left( \langle \pi^+ |(ud)_R d_L |p\rangle_e \right)^2} 
    = |V_{ud}|^2 |(U_\ell)_{1i}|^2 ~,
\end{align}
where in the last equation we use Eq.~\eqref{eq:pi0pipl} and neglect the small difference in the values of $\langle \pi^0 |(ud)_R u_L |p\rangle_{e}$ and $\langle \pi^0 |(ud)_R u_L |p\rangle_{\mu}$. This expression clearly shows that we can determine $|(U_\ell)_{1i}|$ by measuring the ratio $R_{\ell_i}$. By using the unitarity of $U_\ell$, we also obtain the following inequalities: 
\begin{align}
    R_{\ell_i} &\leq |V_{ud}|^2  ~, \label{eq:rliineq} \\ 
    R_e + R_\mu &\leq |V_{ud}|^2 ~,
\end{align}
with $|V_{ud}|^2 \simeq 0.95$~\cite{ParticleDataGroup:2022pth}. These inequalities are useful to discriminate the GUT models. For example, if the nucleon decay operators are induced predominantly by the SU(5) gauge boson exchange in the standard SU(5), $R_e$ is predicted to be~\cite{Ellis:2020qad}\footnote{The gauge boson exchange in the standard SU(5) induces two-types of dimension-six effective operators. The renormalization factors for these operators are slightly different; in Eq.~\eqref{eq:resu5}, we neglect this difference, which turns out to be a good approximation for a typical low-energy mass spectrum~\cite{Ellis:2020qad}.} 
\begin{align}
    R_e|_{\mathrm{SU}(5)} = \frac{1 + (1 + |V_{ud}|^2)^2}{|V_{ud}|^2} \simeq 5 ~,
    \label{eq:resu5}
\end{align} 
which clearly violates the above inequalities. This observation indicates that the measurement of the lifetimes of only two decay channels,  $p \to \pi^0 e^+$ and $n \to \pi^0 \bar{\nu}$, is already capable of distinguishing this scenario from flipped SU(5). Let us also show $R_\mu$ in this case just for completeness~\cite{Ellis:2020qad}: 
\begin{equation}
    R_\mu|_{\mathrm{SU}(5)} = |V_{us}|^2 \simeq 0.05 ~,
\end{equation}
which is consistent with the condition~\eqref{eq:rliineq}. 

As discussed in Ref.~\cite{Ellis:2020qad}, in the flipped SU(5) scenario considered in Refs.~\cite{Ellis:2019jha, Ellis:2019opr, Ellis:2020lnc}, the matrix elements $(U_\ell)_{11}$ and $(U_\ell)_{12}$ are given by the PMNS matrix elements as 
\begin{align}
    (U_\ell)_{11} &= 
    \begin{cases}
        U_{11} & \text{for normal order} \\ 
        U_{13} & \text{for inverted order}
    \end{cases}
    \nonumber
    \\ 
    (U_\ell)_{12} &= 
    \begin{cases}
        U_{21} & \text{for normal order} \\ 
        U_{23} & \text{for inverted order}
    \end{cases}
    ~.
    \label{eq:ulpmns}
\end{align}
These relations hold also in the case where the neutrino mass matrix $m_\nu$ in Eq.~\eqref{eq:mnud} is almost diagonal. We can test this prediction through the measurement of $R_{\ell_i}$.  

To see the dependence of $R_{\ell_i}$ on the unknown matrix elements $(U_\ell)_{11}$ and $(U_\ell)_{12}$, it is convenient to parametrize them by means of two real parameters, $\xi$ and $\phi$. The unitarity condition, 
\begin{equation}
    |(U_\ell)_{11}|^2 + |(U_\ell)_{12}|^2 + |(U_\ell)_{13}|^2 = 1 ~,
\end{equation} 
indicates that it is possible to parametrize these matrix elements as\footnote{It is possible to extend this parametrization to the whole matrix elements with additional seven real parameters. In general, any $3 \times 3$ unitary matrix can be parameterized by three mixing angles and six phases, and we can identify two of the three mixing angles as $\phi$ and $\xi$. For a concrete expression, see, \textit{e.g.}, Ref.~\cite{Rasin:1997pn}. 
} 
\begin{align}
    |(U_\ell)_{11}| = \cos \phi \cos \xi ~, \qquad 
    |(U_\ell)_{12}| = \sin \phi \cos \xi ~, \qquad
    |(U_\ell)_{13}| = \sin \xi ~, 
\end{align}
with $0 \leq \phi, \xi \leq \pi/2$. 

\begin{figure}
    \centering
    \subcaptionbox{\label{fig:Re}
    $R_e$ 
    }
    {\includegraphics[width=0.48\textwidth]{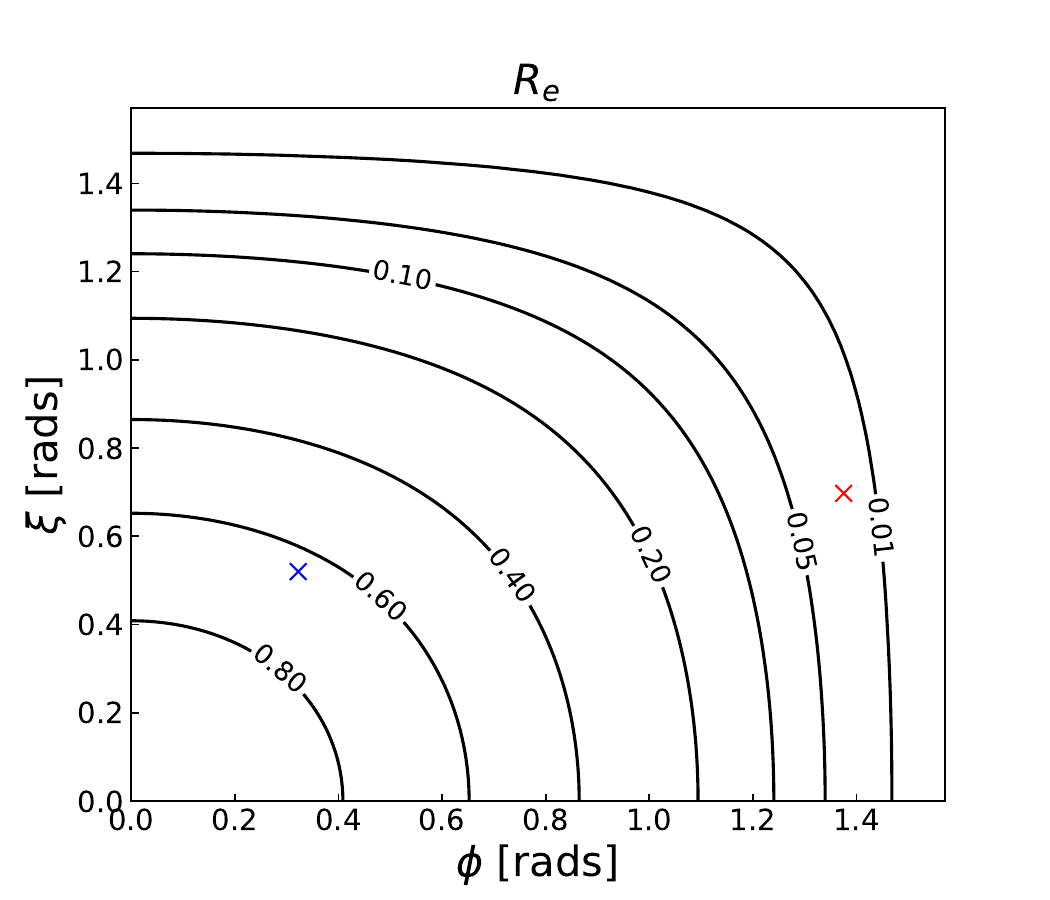}}
    \subcaptionbox{\label{fig:Rmu}
    $R_{\mu}$ 
    }
    { 
    \includegraphics[width=0.48\textwidth]{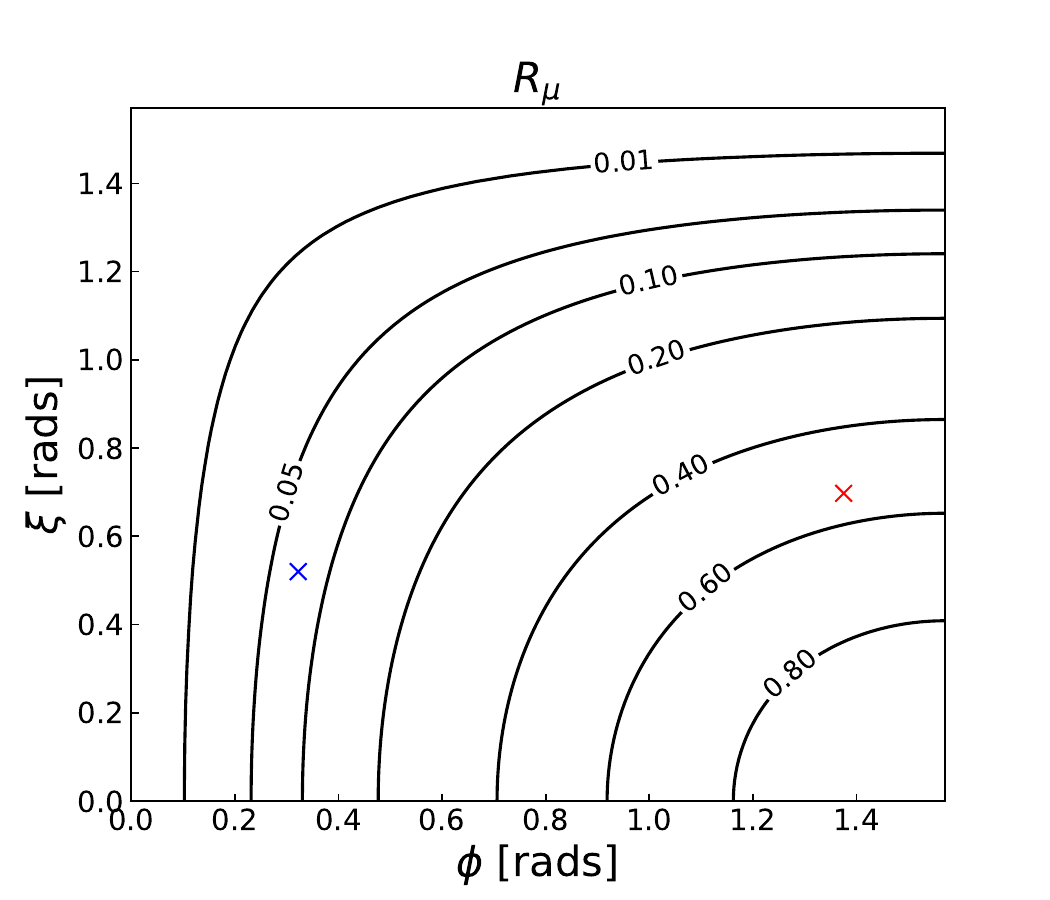}}
    \caption{Contour plots of (a) $R_e$ and (b) $R_\mu$ in the $\phi$-$\xi$ plane. The blue and red crosses correspond to the choice in Eq.~\eqref{eq:ulpmns} with the normal and inverted orderings, respectively. 
    \label{fig:ratio}}
    \end{figure}  

In Fig.~\ref{fig:ratio}, we show the contour plots of (a) $R_e$ and (b) $R_\mu$ in the $\phi$-$\xi$ plane. The blue and red crosses correspond to the choice in Eq.~\eqref{eq:ulpmns} with the normal and inverted orderings, respectively, which is the case for the scenario considered in Refs.~\cite{Ellis:2019jha, Ellis:2019opr, Ellis:2020lnc}. We use \texttt{NuFIT 5.3}~\cite{nufit} to compute the PMNS matrix elements.\footnote{It is found that $|(U_\ell)_{1i}|$ is independent of the unknown Majorana phases in the PMNS matrix. } As we see, $R_e$ and $R_\mu$ play complementary roles in probing the parameter space. It is also found that both $R_e$ and $R_\mu$ are suppressed for $\xi \simeq \pi/2$; in this case, $p \to \pi^+ \bar{\nu}$ and $n \to \pi^0 \bar{\nu}$ become the dominant channel for proton and neutron decays, respectively. This pattern of nucleon decay is distinctive. In the minimal SUSY SU(5), for instance, the decay rate of $p \to \pi^+ \bar{\nu}$ can be much larger than those of $p \to \pi^0 \ell^+$ if the contribution of the dimension-five operators dominates the dimension-six one (see, e.g., Refs.~\cite{Nagata:2013sba, Ellis:2019fwf}). However, in this case, the rate of the $p \to K^+ \bar{\nu}$ decay channel also increases, whilst this does not occur in flipped SU(5). We can thus distinguish flipped SU(5) from the standard SU(5) even for small $\xi$, by looking into the $p \to \pi^+ \bar{\nu}$ (or $n \to \pi^0 \bar{\nu}$) and $p \to K^+ \bar{\nu}$ channels.

\section{Conclusions and discussion}

We have examined the pattern of nucleon decay branching fractions in flipped SU(5). Since nucleon decay is induced by the gauge interactions in flipped SU(5), we can make relatively robust predictions for certain branching fractions, such as $\Gamma(p \to K^0 e^+)/\Gamma(p \to \pi^0 e^+)$ and $\Gamma(p \to K^0 \mu^+)/\Gamma(p \to \pi^0 \mu^+)$. Another notable prediction is the suppression of the $p \to K^+ \bar{\nu}$ channel. However, the calculation of some decay channels is affected by uncertainties due to an unknown unitary matrix $U_\ell$. We have shown that this uncertainty can be parametrized using two real parameters, $\phi$ and $\xi$. These parameters can be determined by measuring the ratios $R_e$ and $R_\mu$, as illustrated in Fig.~\ref{fig:ratio}. Additionally, we have derived upper limits on these ratios, which can be useful for testing flipped SU(5) in future nucleon decay experiments.

Extended scenarios for flipped SU(5), as discussed in Refs.~\cite{Hamaguchi:2020tet, Mehmood:2020irm, Haba:2021rzs, King:2023wkm}, could present different patterns of nucleon decay branching fractions than those considered in this paper. The model dependence is encapsulated in the Wilson coefficients of the nucleon-decay effective operators at low energies, with their chirality and flavor structures influencing the nucleon decay channels. By measuring nucleon decays, we can examine these structures, allowing us to distinguish between different GUT scenarios. A detailed and comprehensive study on this topic will be presented in future work~\cite{HHNT}.

\vspace{6pt}

 \funding{This work was supported by JSPS KAKENHI Grant Numbers 24H02244 (KH), 24K07041 (KH), 21K13916 (NN), 22KJ1022 (SH), and 24KJ0913 (HT).}

\begin{adjustwidth}{-\extralength}{0cm}

\reftitle{References}



\bibliography{ref}

\PublishersNote{}
\end{adjustwidth}
\end{document}